# Creativity as Variations on a Theme:

# Formalizations, Evidence, and Engineered Applications

Rohan Agarwal

Georgia Institute of Technology

roaga@gatech.edu

**Abstract**

There are many philosophies and theories on what creativity is and how it works, but one popular idea is that of variations on a theme and intersection of concepts. This literature review explores philosophical proposals of how creativity emerges from variations on a theme, and how formalizations of these proposals in human subject studies and computational methods result in creativity. Specifically, the philosophical idea of intangible clouds of concepts is analyzed with empirical studies of concept representation and mental model formation, and mathematical formalizations of such ideas. Empirical findings on emergent neural activity from neural network combinations are also examined for evidence of novel, emergent ideas from the collision of existing ones. Finally, work on human-AI co-creativity is used as a lens for concept collision and the effectiveness of this model of creativity. This paper also proposes directions for further research in studying creativity as variations on a theme.

**Introduction**

Creativity is often said to be one of the hallmarks that make us human, and it is a trait that serves us every day, from art and music to innovative technology and problem-solving. However, it is not well understood how it works; how does the human brain be creative so effectively? There are many philosophical ideas on how creativity should be modeled cognitively, but this paper examines the popular proposal that creativity results from variations on a theme. Rather than effortful creation, this model imagines existing concepts subconsciously colliding and overlapping into new, creative ones that give us the creativity we see. There is plenty of anecdotal and philosophical reasoning that supports this idea, as Hofstadter (1982) originally presented. It also serves as a valuable and unique way to think about creativity, originality, and innovation.

However, more formalization and empirical evidence are needed to accept it as a model of creative cognition. To do so, this paper first examines the components of this model and how it imagines mental concepts being represented and how they interact. Then, empirical studies, from both computational and psychological approaches, that examine this form of mental concept representation are reviewed. While these studies formalize concept representation, insight into concept collision is still needed. A study using simulated neural network activity is reviewed to provide evidence for this idea. The field of human-AI co-creativity is also explored as an example and potentially useful application of concept collision, and more generally, this model of creativity. It is crucial to understand how this philosophical perspective on creativity can function and benefit us in a more concrete way. Engineered co-creative applications shed light on the empirical effectiveness of this model and its translation to real-world creativity, not just its correctness with respect to human cognition.

**Review**

*Philosophical Proposal*

Hofstadter (1982) set out to answer the fundamental question of how humans can imagine ideas that do not already exist (i.e., to be creative and original), which are notably often sensible and valuable (like an invention, scientific theory, or a piece of a music). He claimed that creativity stems from variations on existing themes and concepts in the creator's mind. He then justified his claim through a series of thought experiments and anecdotes, which while convincing, underscore the need for scientific evidence to support this model of creativity. First, acknowledging that he lacked a scientific model for what a concept is, he imagined a concept as a box with control knobs used to vary different attributes of the concept. By imagining the consequences of that metaphor for musical compositions and $n$-dimensional Rubik's Cubes, he concluded that a concept has an infinite number of knobs which reveal themselves depending on the concepts in the mind of the person. He noted how one person can vary a theme to create an idea, but others can still take that idea and vary it infinitely further (Hofstadter, 1982).

Hofstadter then proposed subconscious slipping and blending between concepts as the source of these variations, changing the mental context and revealing different sets of knobs to the mind. He modeled concepts as vague clouds representing the possible set of variations and slipping as the intersection of multiple clouds. It is also notable that he advocated for using computers to help humans explore these possibility spaces, which he termed "implicospheres" (Hofstadter, 1982). In summary, the model of creativity proposed here is that of subconscious possibility spaces around concepts, built through manipulatable, context-dependent parameters, that intersect to yield new variations and concepts. Hofstadter, however, provides no scientific basis for this representation of concepts or their blending, so formal studies are now presented.

*Scientific Basis for Variations on a Theme as Creativity*

  Vosniadou (2002) investigated mental models in concept development and reasoning in children, arguing they are a core part of human cognition. She first analyzed their construction by having children draw and answer generative questions about their model of the Earth. Through further generative questioning (e.g., where would you end up if you walked forever?), the children were manipulating their models with imaginary scenarios to answer the questions. It was found that explicit knowledge can integrate with the conceptual model to create novel explanations and further conceptual knowledge. In fact, this is done over time as new information creates new theories in the mind and lets mental models be tweaked and revised. Vosniadou also presented other interviews with adults that highlight how when faced with inconsistencies in their mental model, people creatively revise their model to solve the problem (often leading to misconceptions). The existing mental model in mind determines how new information is interpreted and conceptualized (Vosniadou, 2002). These findings support the idea of iteratively tweaking and growing concepts, and the power of concepts in generating novel creative ideas, especially when trying to integrate new information.

  However, this does not answer how mental models are represented in the brain. Two main theories of concepts are the prototype theory, where categories are defined based on sets of characteristic features, and the exemplar theory, where categories are defined by encountered instances of the category (Murphy, 2016). Murphy (2016) conducted a literature review on both theories and psychological findings surrounding concepts. While there is strong evidence for salient exemplars influencing categorization, he found lackluster evidence on recalling specific exemplars to decide categorization. Murphy (2016) highlights several phenomena that have prototype explanations but no exemplar explanation, such as hierarchical structure of concepts, knowledge effects, and induction. In fact, since exemplar models in the literature seemed to only apply to category learning, and not concepts (i.e., mental models that can be used for learning, communication, and reasoning, as Vosniadou (2002) demonstrated), he argued that there is no exemplar theory of concepts at all (Murphy, 2016). If the prototype theory, or representing concepts through their features, is more accurate to human cognition, then that aligns well with the proposal of variations on a theme. It implies that concepts really are built around a central set of attributes that, if adjusted, can yield new variations of a concept and even new concepts entirely.

Such a model can also be mathematically formalized. Based on the idea of the implicosphere (Hofstadter, 1982), researchers designed a spatial knowledge representation based on projecting the implicosphere onto three planes: the real, the conceptual, and the symbolic (de Mello and de Carvalho, 2015). A cultural filter, or person-dependent context as Hofstadter (1982) described it, was used to transfer real-world phenomena to the conceptual plane, creating different variations of concepts for different people. The authors also demonstrated how this three-plane model can be used to explain reification (imagining new ideas from abstracted concepts), intuition, formalization, interpretation, and deduction of concepts (de Mello and de Carvalho, 2015). This work formalizes the idea of variations on a theme and demonstrates its usefulness for modeling many creative and cognitive tasks, which include key abilities found to apply to mental models (Vosniadou, 2002) and the prototype theory of concepts (Murphy, 2016). This formalization shows that this model of creativity can align effectively with empirical findings on the subject.

More than just concept representation, though, how concepts collide to cause creative output is important to this theory of creativity. While concept combination has been successfully shown with symbolic representations in the past, Thagard and Stewart (2010) demonstrated its more generalizable potential with neural activity—in particular, by using convolution to combine vectors represented neural activity patterns. Their neural representations, more than symbolic/verbal ones, facilitated the fluid creative thinking Hofstadter described (Thagard and Stewart, 2010). The authors first devised a multi-dimensional vector representation of neural activity by multiplying the number of neurons to handle their stochasticity. Then they successfully convolved multiple representations of simulated neural activity into new ones, including patterns that correspond to the heightened emotions upon a creative discovery (e.g., an "aha!" or "eureka!" moment). These simulations provided evidence for the intersection of multimodal concepts in the brain leading to original ideas and a sense of creative discovery. The authors also stated that not every combination will be considered "creative," as Hofstadter (1982) also pointed out for the collision of implicospheres. However, the ones accompanied by the emotional experience of discovery likely are what we consider creative (Thagard and Stewart, 2010). This study directly supports the proposal that the collision of concepts in the brain is the source of creativity, not only in the outputted ideas, but also in the sense in which humans emotionally experience it.

### *Co-Creativity as a Lens*

While all these studies provide great insight into the plausibility of this model of creativity, it is also valuable to consider concrete examples of how it enables creativity. Yannakakis et. al. (2014) introduced the paradigm of mixed-initiative co-creativity (MI-CC): when a human and a computer both proactively collaborate on a creative work, like a story or a video game level. Reframing parts of the possibility space and introducing random stimulus are found to foster the creative process and creative output. The user of a co-creative system often has a diagram or user interface that serves as an extension of their mind, and facilitates creative, lateral thinking (Yannakakis, 2014). Effectively, the AI agent's concept space and the user's concept space collide and interact on this diagram; later works even outline multiple dimensions for such interactions, as if in 3D space, and modularize the different parts of this extended mind (Lin et. al, 2022). Yannakakis et. al. (2014) argued that this interaction (usually suggestions and iterative co-creation) facilitates greater exploration of the possibility space, leading to more novel and valuable creative output. Using a human subject study with a co-creative level design tool called Sentient Sketchbook, analysis of usage of the tool led to the conclusion that MI-CC is useful to human users and strongly guides the creative paths they take (Yannakakis, 2014). While this study is not about creative within the human mind, it demonstrates that when the human mind is effectively extended onto an external diagram, introducing new context, suggesting new ideas and concepts, and exploring the possibility space or variations around a concept lead to superior creative process and output. It is an engineered example that shows that the variations on a theme model is in fact an effective and valuable source of creativity.

      In addition, MI-CC takes great advantage of "control knobs" similar to Hofstadter's (1982) metaphor for varying the features on a concept. For example, Lin and Riedl (2021) introduced a control knob system for guiding the story generation of a large language model. By providing a "sketch," a human user can guide the topics of a generated story, sentence by sentence. Not only does this framework allow for control knobs, but also blending of topics and concepts in the generation. Blending fluency and control fidelity were found to be effective through computational analysis, and a human subject study confirmed preference for the generations from this system. The authors envisioned their system being used for co-creative

applications, as it was in Lin et. al. (2022). The system is remarkably similar to Hofstadter's (1982) metaphor of control knobs on a black box and blending of concepts to generate creative output. The study proves the possibility and effectiveness of a system based on these proposals.

## Discussion

### *Summary*

Variations on a theme as the source of creativity is a common model of how human creativity works, but without empirical evidence, it remained an anecdotal and philosophical one. This paper presented studies that do provide substantial support for this theory, as originally proposed by Hofstadter (1982). Research into how we form and use mental models (Vosniadou, 2002) showed how they can grow, be manipulated, and used to produce novel, creative ideas in response to new context. Further literature review lent support to the prototype theory of concepts (Murphy, 2016), meaning we form concepts around characteristic features which can vary, similar to Hofstadter's idea of implicospheres (Hofstadter, 1982). While Hofstadter's (1982) theory may seem vague and unscientific, implicospheres and changing context can be used as a basis for geometric knowledge representations that explain core creative processes (de Mello and de Carvalho, 2015).

With evidence for variable features forming concepts and leading to creativity, there is also evidence for the intersection of concepts and its benefits for creativity. Thagard and Stewart (2010) demonstrated that multiple neural activity patterns, representing concepts, can be fluidly interweaved into novel activity and lead to the emotional "aha!" moment we often get upon a creative discovery. The field of mixed-initiative co-creativity also lends great support to the creative usefulness of this idea. Yannakakis et. al. (2014) demonstrated that when humans and

AI proactively collaborate on creative tasks, effectively extending the mind of the human and introducing new context, the possible variations are better explored, and creativity is fostered. This paradigm also makes extensive use of control knobs and blending concepts like what Hofstadter (1982) proposed, as shown by Lin and Riedl's (2021) controllable story generation model. Overall, there is evidence in diverse areas of scientific literature for variations on a theme being the source of human creativity, and an effective method for achieving creativity in engineered applications, including creative AI.

### *Limitations and Future Research*

While there is substantial support, the question of the nature of human creativity is not settled. All the literature reviewed has only been able to answer that question through indirect observation, because it is near impossible to parse the inner workings of the brain otherwise. There is still no clear-cut answer to how creativity works, and other theories could also be supported with the same studies. Additionally, all the studies that involve creative output--whether learning how the Earth works, new (but illegible) neural patterns, or video game levels--are very small examples of creativity compared to the most grand and valuable examples of scientific discovery, artistry, or innovation. In case creative cognition changes depending on the quality or scale of creative output, future work should aim to produce more valuable creative works or capture insights from those who do (perhaps through a longitudinal study). Some people are also more creative personality-wise, more successfully creative, or have different creative processes and styles. Further investigation is also needed in identifying how cognition varies between people—can one theory apply to everyone and everything we consider creative?